\documentclass[%
superscriptaddress,
twocolumn,
showpacs,preprintnumbers,
amsmath,amssymb,
aps,
]{revtex4-1}

\usepackage{graphicx}
\usepackage{dcolumn}
\usepackage{bm}
\usepackage[mathlines]{lineno}
\usepackage{hyperref}
\hypersetup{
    colorlinks=true,
    linkcolor=blue,
    filecolor=blue,
    urlcolor=blue,
    citecolor=blue,
}

\begin{document}

\title{CrSbSe$_{3}$: a pseudo one-dimensional ferromagnetic semiconductor}

\author{Guangyu Wang}
\affiliation{Laboratory for Computational Physical Sciences (MOE),
	State Key Laboratory of Surface Physics, and Department of Physics,
	Fudan University, Shanghai 200433, China}
\affiliation{Shanghai Qi Zhi Institute, Shanghai 200232, China}

\author{Lu Liu}
\affiliation{Laboratory for Computational Physical Sciences (MOE),
	State Key Laboratory of Surface Physics, and Department of Physics,
	Fudan University, Shanghai 200433, China}
\affiliation{Shanghai Qi Zhi Institute, Shanghai 200232, China}

\author{Ke Yang}
\affiliation{College of Science, University of Shanghai for Science and Technology,
       Shanghai 200093, China}
\affiliation{Laboratory for Computational Physical Sciences (MOE),
	State Key Laboratory of Surface Physics, and Department of Physics,
	Fudan University, Shanghai 200433, China}

\author{Hua Wu}
\email{Corresponding author: wuh@fudan.edu.cn}
\affiliation{Laboratory for Computational Physical Sciences (MOE),
	State Key Laboratory of Surface Physics, and Department of Physics,
	Fudan University, Shanghai 200433, China}
\affiliation{Shanghai Qi Zhi Institute, Shanghai 200232, China}
\affiliation{Collaborative Innovation Center of Advanced Microstructures,
	Nanjing 210093, China}


\begin{abstract}
Low-dimensional magnetic materials have attracted much attention due to their novel properties and high potential for spintronic applications. In this work, we study the electronic structure and magnetic properties of the pseudo one-dimensional compound CrSbSe$_3$, using density functional calculations, superexchange model analyses, and Monte Carlo simulations. We find that CrSbSe$_3$ is a ferromagnetic (FM) semiconductor with a band gap of about 0.65 eV.
The FM couplings within each zig-zag spin chain are due to the Cr-Se-Cr superexchange with the near-90$^\circ$ bonds, and the inter-chain FM couplings are one order of magnitude weaker. By inclusion of the spin-orbit coupling (SOC) effects, our calculations reproduce the experimental observation of the easy magnetization $a$ axis and the hard $b$ axis (the spin-chain direction), with the calculated moderate magnetic anisotropy of 0.19 meV/Cr. Moreover, we identify the nearly equal contributions from the single ion anisotropy of Cr, and from the exchange anisotropy due to the strong SOC of the heavy elements Sb and Se and their couplings with Cr. Using the parameters of the magnetic coupling and anisotropy from the above calculations, our Monte Carlo simulations yield the Curie temperature $T_{\rm{C}}$ of 108 K.
\end{abstract}

\maketitle

\section{INTRODUCTION}
Low-dimensional magnetic systems attract widespread attention for their novel physical properties and practical applications.
Recently, two-dimensional (2D) ferromagnetism (FM) has been observed in the atomically thin layers CrI$_{3}$ and  Cr$_{2}$Ge$_{2}$Te$_{6}$~\cite{Huang_CrI3, Gong_CrGeTe}, marking the beginning of a new chapter in 2D magnetism.
The development of 2D magnetic materials would greatly accelerate the size reduction and promote efficiency in postsilicon microelectronics~\cite{Burch2018, cm2015}. As the one-dimensional (1D) and 2D isotropic Heisenberg exchange systems do not have long-range magnetic order at finite temperatures, according to the Mermin-Wagner theorem~\cite{MW}, the emergent magnetism in low-dimensional materials should be accompanied by a magnetic anisotropy (MA).
Compared to the 2D magnetic materials, a magnetic ordering is even harder to establish in 1D systems since the fluctuations become much more aggressive~\cite{Landau, Lecture_Mag}.
Naturally, an experimental realization of 1D FM is more challenging.

\begin{figure}[t]
	\includegraphics[width=7.7cm]{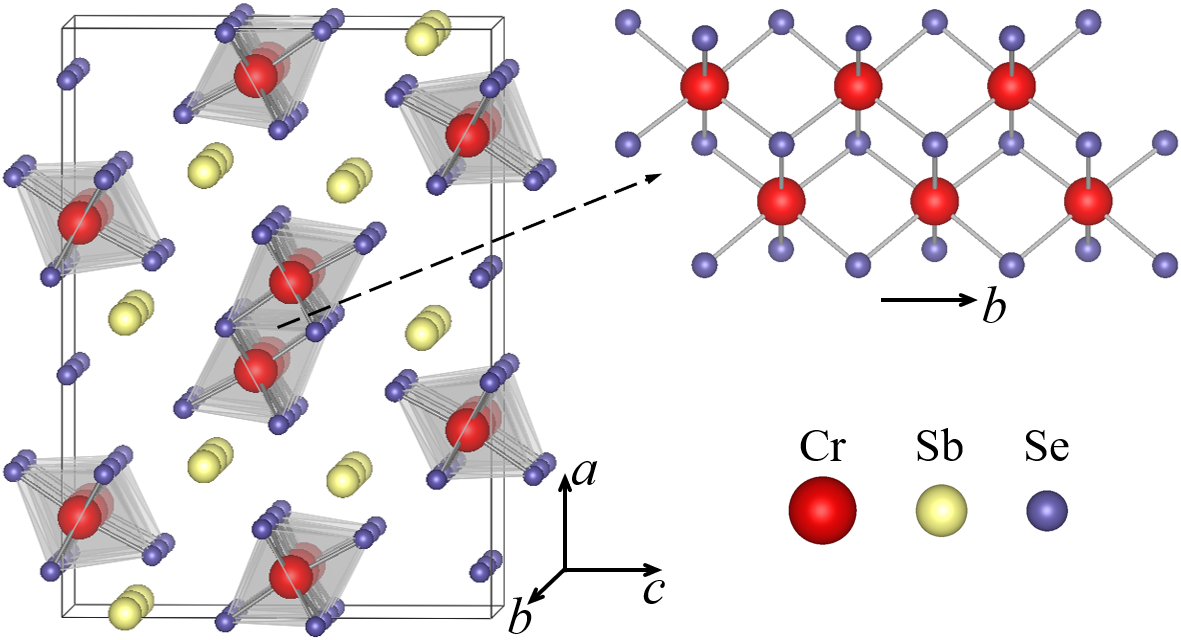}
	\caption {The $2\times2\times1$ supercell in the pseudo-1D structure of CrSbSe$_3$ with the double rutile chains formed by the edge-sharing CrSe$_6$ octahedra, i.e., the zig-zag Cr-spin chains along the $b$ axis.}
	\label{fig1}
\end{figure}

Ternary chromium trichalcogenides are a platform for low-dimensional magnets, and among them, Cr(Si, Ge)Te$_3$ are 2D materials of the van der Waals type, with the CrTe$_{6}$ octahedra forming a honeycomb-layered structure~\cite{Gong_CrGeTe, Zhang_PRL2019}.
On the other hand, Cr(Sb, Ga)X$_{3}$ (X=S, Se) possess a pseudo-1D structure, in which CrX$_{6}$ octahedra form the edge-sharing double rutile chains (the zig-zag Cr-spin chains) along the $b$ axis and Sb or Ga atoms linking neighboring chains~\cite{Odink1993, Volkov1997}, see Fig.~\ref{fig1}. Very recently, the pseudo-1D CrSbSe$_{3}$ received a revived interest. It is a FM semiconductor with the Curie temperature of 71 K, the band gap of 0.7 eV, and the easy magnetization $a$ axis and hard $b$ axis~\cite{PRM2018, Sun2020, PhysRevB2020}.
A recent work presented some numerical results of the structural details and a very brief description of the electronic structure and magnetism of CrSbSe$_3$ (and CrSbS$_3$), but no physical picture was discussed therein~\cite{mathew2020}.
Note that in most cases, a FM material is metallic while an antiferromagnetic (AFM) one is insulating.
Thus, either a FM insulator/semiconductor or an AFM metal seems to be an exception and would be of interest, let alone the low-dimensional magnetic order.

\begin{figure}[t]
	\includegraphics[width=7cm]{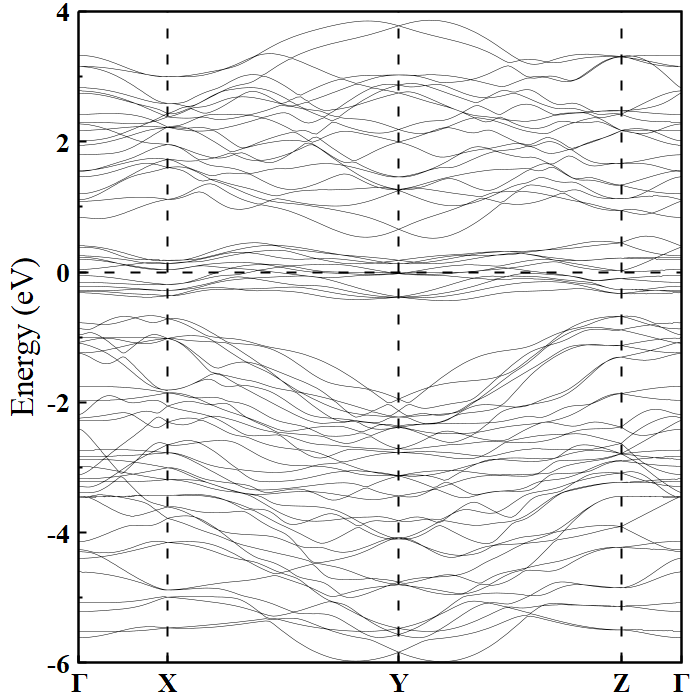}
	\caption {Band structure of CrSbSe$_3$ by LDA: a low-dimensional narrow band system.
    Fermi level is set at zero energy.
	}
    \label{fig2}
\end{figure}

In this work, we study the electronic structure and magnetism of CrSbSe$_{3}$, using density functional calculations, superexchange model analyses, and Monte Carlo simulations. Our work finds the strongly covalent semiconducting behavior and determines the FM ground state with the easy magnetization $a$ axis and hard $b$ axis. The intra-chain FM couplings are rationalized by the superexchange of the near-90$^\circ$ Cr-Se-Cr bonds (with the Se $4p$ double holes intermediate state) within the pseudo-1D zig-zag spin chains, and the inter-chain FM couplings are one order of magnitude weaker. Moreover, we identify the single ion anisotropy of Cr and the exchange anisotropy due to the SOC of Sb $5p$ and Se $4p$ and their coupling with Cr $3d$, and both the anisotropies are comparably moderate, being about 0.1 meV/Cr. Using the parameters of the magnetic couplings and anisotropies, our Monte Carlo simulations account for the experimental $T_{\rm C}$.

\section{COMPUTATIONAL DETAILS}

Density functional calculations are carried out using Vienna $ab$ $initio$ Simulation Package with the local-spin-density approximation (LSDA)~\cite{VASP}.
The projector augmented-wave method with a plane-wave basis set is used and the cut-off energy is set to 400 eV~\cite{PAW}.
We use the experimental lattice constants of CrSbSe$_3$ measured by single-crystal x-ray diffraction at 296 K~\cite{PRM2018}, and carry out atomic relaxation until each atomic force is less than 0.01 eV/\AA~(the resultant Cr-Se-Cr bond lengths and angles turn out to be almost the same (within 0.4\%) as the experimental ones). A $2 \times 2 \times 1$ supercell is used, and it accommodates all the FM and AFM structures we study below. The LSDA plus Hubbard $U$ (LSDA+$U$) method~\cite{LSDAU1, LSDAU2} is used to describe the correlation effect of the Cr $3d$ electrons, with a common value of $U$=3 eV and Hund exchange $J\rm_{H}$=1 eV. Note that typically, $U$ is about 3 eV for low-dimensional Cr-compounds~\cite{guoyilv_2018, chens_2020, wanghua_2020}. $J\rm_{H}$ is actually the difference of the energies of electrons with different spins or orbitals on a same atomic shell, and therefore, $J\rm_{H}$ is almost not screened and not modified when going from an atom to a solid. It is almost a constant for a given element and is typically 0.8-1.0 eV for a 3$d$ transition metal~\cite{Khomskii_2014}. As seen below, our LSDA+$U$ calculations well reproduce the experimental semiconducting gap.
The Monkhorst-Pack $k$-mesh of $3 \times 3 \times 2$ is sampled for integration over the Brillouin zone~\cite{K_Mesh_MP}, and the $5 \times 5 \times 3$ $k$-mesh is also tested, both showing practically the same results and ensuring a sufficient accuracy.
We also include the spin-orbit coupling (SOC) in our LSDA+$U$+SOC calculations to study the magnetic anisotropy.
Moreover, we carry out Monte Carlo simulations to estimate the Curie temperature, and the simulation details and results will be seen below.

\section{RESULTS AND DISCUSSION}

\begin{figure}[t]
	\includegraphics[width=8cm]{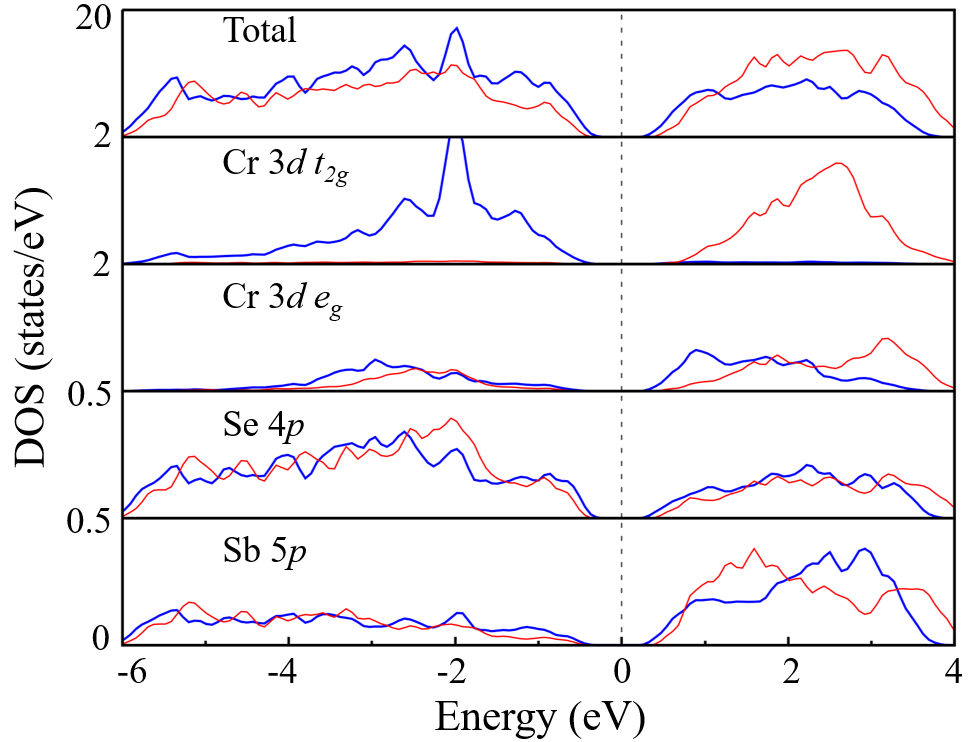}
	\caption {Density of states (DOS) of CrSbSe$_3$ by LSDA+$U$: a strongly covalent Mott type semiconductor.
    The blue (red) curves stand for the up (down) spin channel.
    Fermi level is set at zero energy.
	}
    \label{fig3}
\end{figure}

We first carry out the spin-constrained LDA calculation, and the obtained band structure is shown in Fig.~\ref{fig2}. Apparently, CrSbSe$_{3}$ is a low-dimensional narrow band system. In particular, a group of bands around the Fermi level arise mainly from the $t_{2g}$ states of the Cr $3d$ orbitals, and they have the bandwidth less than 1 eV. Naturally, one expects the Coulomb correlation effects of the localized Cr $3d$ electrons.

We then perform  the LSDA$+U$ calculation to study the electronic structure of CrSbSe$_{3}$. Fig.~\ref{fig3} shows the total and orbitally resolved density-of-states (DOS) results in the FM state.
We find that CrSbSe$_3$ has a semiconducting band gap of 0.65 eV and it is in a good agreement with the experimental one of 0.7 eV~\cite{PRM2018}. As the $U$ value (here $U$= 3 eV) is much bigger than the bandwidth (less than 1 eV) of the Cr $3d$ electrons, one can name CrSbSe$_3$ a Mott type semiconductor. The Cr 3$d$ orbitals in the local CrSe$_6$ octahedral crystal field split into the lower $t_{2g}$ triplet and higher $e_g$ doublet. The up-spin $t_{2g}$ orbitals are fully occupied, and the down-spin $t_{2g}$ are empty. The higher $e_g$ orbitals are formally unoccupied, but they seem to be partially occupied due to the strong covalence with the occupied Se $4p$ orbitals. Therefore, here the Cr ion is in the formal 3+ state and has the formal $t_{2g}^3$ configuration with the local spin=3/2. The integer total spin moment of 3 $\mu_{\rm B}$ per formula unit (fu) in this FM semiconducting solution agrees well with the experiments~\cite{PRM2018, Sun2020, PhysRevB2020}. As the Se $4p$ and Sb $5p$ orbitals are fat ones and they have strong covalence with the Cr $3d$ orbitals, we can call CrSbSe$_{3}$ a strongly covalent Mott type semiconductor~\cite{Zaanen_PRL1985, Khomskii_2014}.
Moreover, it is the Cr-Se covalence that makes the formal Se$^{2-}$ ion have some unoccupied $4p$ components in the energy window of the unoccupied Cr $3d$ states.
For this reason, the fat Se $4p$ orbitals get a minor negative spin polarization, as seen in Table~\ref{tb1}. The even fatter Sb $5p$ orbitals are also slightly spin-polarized.
Those strong covalences are important to the magnetic couplings and anisotropies as studied below.

\renewcommand\arraystretch{1.3}
\begin{table}[t]
\footnotesize
	\setlength{\tabcolsep}{1.8mm}
	\caption{
    Relative total energies $\Delta E$ (meV/fu), total spin moments $M_{\rm{tot}}$ ($\mu_{\rm{B}}$/fu) and local spin moments ($\mu_{\rm{B}}$) of CrSbSe$_3$ in different magnetic states with the $a$-axis magnetization by LSDA+$U$+SOC. The derived exchange parameters (meV) are given, see also Fig.~\ref{fig4} (note that the far distanced intra-chain $J$'=$-$0.06 meV is two orders of magnitude weaker than $J_1$ and $J_2$ and thus negligible). The results are also included for the FM ground state with the $b$- or $c$-axis magnetization.
	}
	\label{tb1}
	\begin{tabular}{p{0.55cm}<{\centering} p{1.2cm}<{\raggedleft}  p{1.2cm}<{\raggedleft} p{1.2cm}<{\raggedleft} p{1.2cm}<{\raggedleft} p{1.2cm}<{\raggedleft} }
		\hline\hline
                    States  & $\Delta E$ &$M_{\rm{tot}}$             & Cr                      & Sb                        & Se \\ \hline
              FM$^a$    &   0.00          &3.00          & 2.98                  & $-$0.03                &  $-$0.04   \\
		               AFM1   &  35.85        &0.00          & $\pm$2.88       & 0.00                     & $\mp$0.02\\
		               AFM2   &  26.43        &0.00          & $\pm$2.89       & $\mp$0.03          & $\mp$0.03\\
                       AFM3   &   0.57         &0.00          & $\pm$2.98       & $\mp$0.02          & $\mp$0.04\\
                       AFM4   &   2.26         &0.00          & $\pm$2.97       & $\mp$0.03          & $\mp$0.03\\  \hline
                       FM$^b$   &   0.19         &3.00           & 2.98                  & $-$0.03                &  $-$0.04   \\
               FM$^c$   &   0.02         &3.00           & 2.98                  & $-$0.03                &  $-$0.04   \\ \hline \hline
                             &$J_{1}$      & $J_{2}$             & $J_{3}$                & $J_{4}$ &   \\
		                    &$-$5.56      & $-$4.88            &$-$0.26                & $-$0.50 &   \\  \hline \hline
	\end{tabular}
\end{table}

\begin{figure}[t]
	\includegraphics[width=8cm]{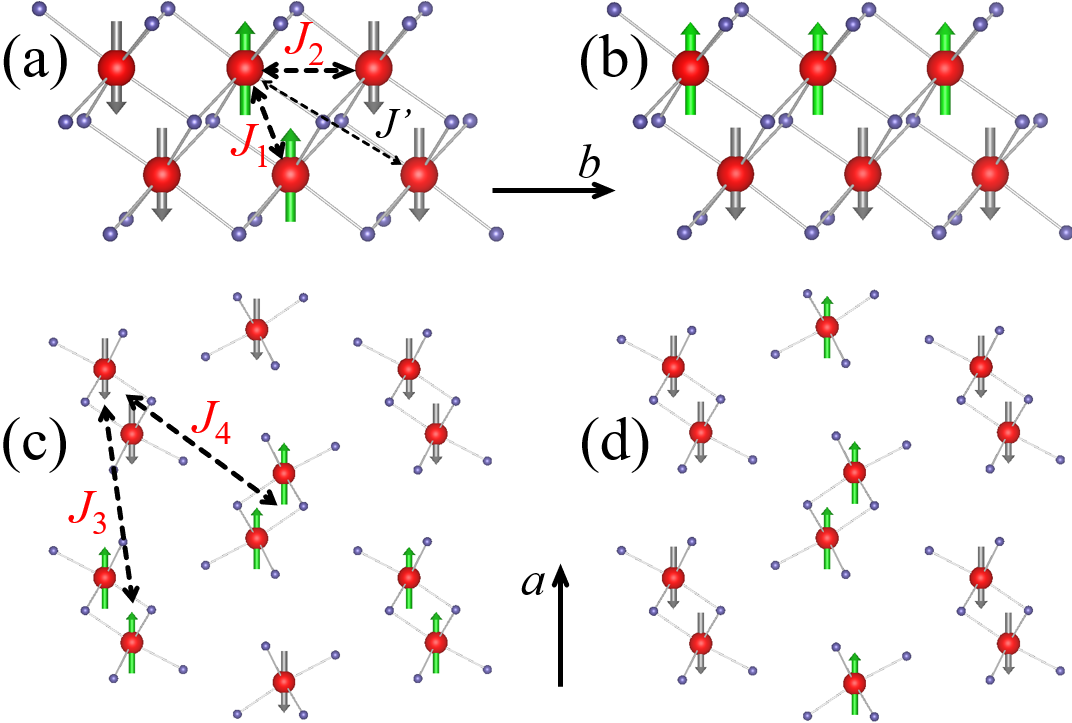}
	\caption {The AFM1/2/3/4 structures consecutively from (a) to (d). $J_1$ and $J_2$ are the exchange interactions within the zig-zag chain, and $J_3$ and $J_4$ are the inter-chain ones. The far distanced intra-chain $J$' turns out to be negligible.}
\label{fig4}
\end{figure}

We now study the magnetic structure of CrSbSe$_3$. To determine the magnetic ground state and the exchange parameters, we choose four different AFM structures (see Fig.~\ref{fig4}), besides the above FM state. The magnetic anisotropies are also of our concern. Therefore, here we include the SOC effects and carry out LSDA+$U$+SOC calculations. We define the intra-chain exchange parameters $J_1$ and $J_2$ in the zig-zag spin chains ($S$=3/2), and the inter-chain $J_3$ and $J_4$, see Fig. \ref{fig4}. Then, counting $JS^2$ for each spin pair, the magnetic exchange energies (per fu) of the FM state and four AFM states are written as follows
\begin{equation*}
\begin{aligned}
E_{\rm{FM}}  &=  (J_1 + J_2 + 0.5J_3 + J_4)S^2, \\
E_{\rm{AFM1}} &= -J_2S^2,\\
E_{\rm{AFM2}} &= (-J_1 + J_2)S^2, \\
E_{\rm{AFM3}} &=  (J_1 + J_2 - 0.5J_3 + J_4)S^2, \\
E_{\rm{AFM4}} &=  (J_1 + J_2 + 0.5J_3 - J_4)S^2.
\end{aligned}
\end{equation*}
As seen from the total energy results in Table~\ref{tb1}, the FM solution is the ground state. Using the values of the relative total energies and the above equations, one can derive the intra-chain $J_1$=$-$5.56 meV and $J_2$=$-$4.88 meV, and the inter-chain $J_3$=$-$0.26 meV and $J_4$=$-$0.50 meV being one order of magnitude weaker. The far distanced intra-chain $J$' is calculated to be $-$0.06 meV, and it is two orders of magnitude weaker than $J_1$ and $J_2$ and thus negligible. All this accords with the pseudo-1D structure of CrSbSe$_3$.
All these negative values indicate that both the intra- and inter-chain couplings are FM.
In the FM ground state, the $S$=3/2 Cr$^{3+}$ ion has the local spin moment of
2.98 $\mu_{\rm B}$, and the formal Sb$^{3+}$ and Se$^{2-}$ ions (both having strong covalence) carry a minor negative spin moment of $-$0.03 and $-$0.04 $\mu_{\rm B}$. Owing to the formally closed Cr$^{3+}$ $t_{2g}^3$ shell and the formal Sb$^{3+}$ 5$p^0$ and Se$^{2-}$ 4$p^6$ configurations, the SOC induced orbital moments are marginally small, being less than 0.01 $\mu_{\rm B}$ for each of the three elements. As a result, the total magnetic (spin) moment of 3 $\mu_{\rm B}$/fu in the FM ground state (including a contribution of 0.17 $\mu_{\rm B}$/fu from the interstitial region) agrees well with the experiments~\cite{PRM2018, Sun2020, PhysRevB2020}.

\begin{figure}[t]
	\includegraphics[width=6.1cm]{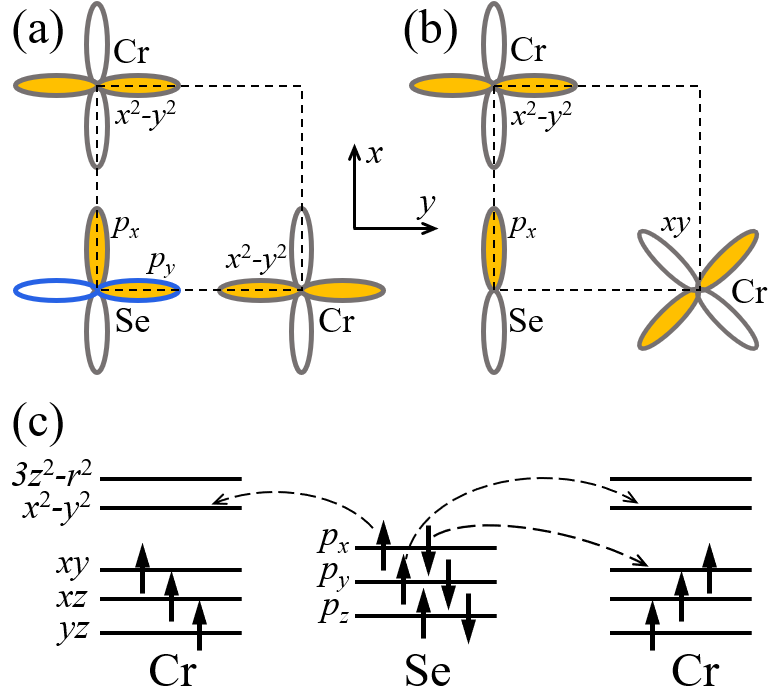}
	\caption {
    (a) and (c): FM superexchange in the double rutile chains via the $(x^{2}-y^{2})$-$(p_{x},p_{y})$-$(x^{2}-y^{2})$ orbitals.
    (b) and (c): FM superexchange via the $(x^{2}-y^{2})-p_{x}-xy$ orbitals.
	}
\label{fig5}
\end{figure}

We now have a close look at the two major intra-chain FM couplings $J_1$ and $J_2$, which occur between two edge-sharing CrSe$_6$ octahedra, see Fig.~\ref{fig4}(a). This situation is similar to the extensively studied 2D material CrI$_3$~\cite{Wang_2016, 2D_Lado2017, Kim_PRL2019}, whose planar honeycomb lattice is composed of the edge-sharing CrI$_6$ octahedra. In the insulating CrSbSe$_3$, the direct Cr-Cr coupling should be AFM, considering the Pauli exclusion principle in the intermediate $t^2_{2g}$-$t^4_{2g}$ state virtually excited from the $t^3_{2g}$-$t^3_{2g}$ ground state. However, the Cr-Cr distances of 3.62~\AA~ for $J_1$ and 3.78~\AA~for $J_2$ are already much larger than that of 2.52~\AA~in the bulk Cr metal, and therefore, the direct AFM exchange should be quite weak. Instead, the near-90$^\circ$ Cr-Se-Cr superexchange should be dominant for both the $J_1$ and $J_2$ via the strongly covalent Cr-Se bonds. Considering the charge-transfer type band gap of CrSbSe$_3$, the most accessible charge excitation is from the Se $4p$ to Cr $3d$. Therefore, the virtually excited intermediate state Cr $3d^4$-Se $4p^4$-Cr $3d^4$ with the $4p$ double holes
should be most plausible. Then we plot in Fig.~\ref{fig5} two major superexchange paths, one via the two strong $pd\sigma$ bonds (Cr $x^2-y^2$ and Se $4p_x\uparrow$) and (Se $4p_y\uparrow$ and Cr $x^2-y^2$),
and the other via one strong $pd\sigma$ bond (Cr $x^2-y^2$ and Se $4p_x\uparrow$) and one moderate $pd \pi$ bond (Se $4p_x\downarrow$ and Cr $xy$). Considering the local Hund exchange and Pauli exclusion principle (see Fig.~\ref{fig5}(c)), both the superexchange paths give the effective Cr-Cr FM couplings within the zig-zag chains.

\begin{figure}[t]
	\includegraphics[width=8.5cm]{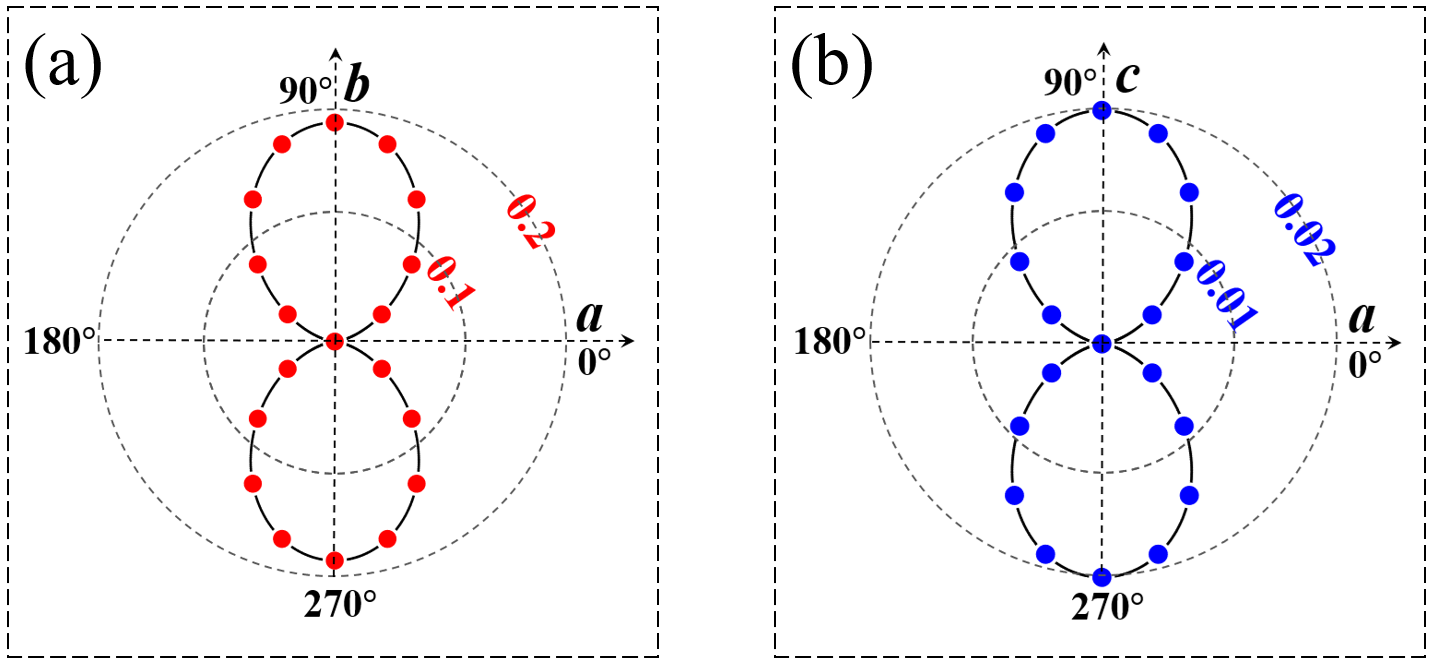}
	\caption {
    Polar diagrams of the MAE in the $ab$ plane (a) and $ac$ plane (b).
    The dashed circles with different radius represent the magnitude of MAE (in unit of meV/Cr).
    The red and blue dots refer to the MAE values obtained in our calculations.
	}
\label{fig6}
\end{figure}

Magnetic anisotropy is of great concern for low-dimensional magnets. Here we carry out LSDA+$U$+SOC calculations by assuming the different magnetization axes within the $ab$ or $ac$ plane (see Fig.~ \ref{fig1}). We plot the polar diagrams of the magnetic anisotropy energy (MAE) in Fig.~\ref{fig6}. Our results represented by the data points show that the $a$ axis is the easy magnetization axis, which is perpendicular to the double rutile chains along the $b$ axis.
The $b$ axis turns out to be the hard one with the MAE of 0.19 meV/Cr, and the $c$ axis is the intermediate one with MAE of 0.02 meV/Cr relative to the easy $a$ axis, see also Table~\ref{tb1}. Our finding of the easy magnetization axis reproduces the experimental observation~\cite{PRM2018, Sun2020}. Moreover, the experimental saturation magnetic fields of 0.1, 1 and 1.5 Tesla respectively along the $a$, $c$ and $b$ axes indicate that the MAE is at the energy scale of 0.1 meV/Cr, being in agreement with our calculations.

Now we seek the origin of the MA which breaks the spin rotational invariance. Single ion anisotropy (SIA) and exchange anisotropy (EA) are quite common in low-dimensional magnets~\cite{2D_Lado2017, Kim_PRL2019, VI3, VBr3JMCC}.
The magnetic behavior of a system with strong SIA may be described by Ising model. For example, VI$_3$ monolayer, out of its bulk van der Waals structure, was recently predicted to carry a large orbital moment and have a strong perpendicular SIA of about 15 meV per V$^{3+}$ ion. In this sense, VI$_3$ monolayer could be a 2D Ising FM insulator~\cite{VI3}.
The SIA arises from the intrinsic SOC effect
$\lambda \overrightarrow{L} \cdot \overrightarrow{S}$, where the orbital momentum must be present (although it is quenched in many cases due to the crystal field and band formation, $etc$). In CrSbSe$_3$, the formal Cr$^{3+}$ ion has the closed $t_{2g}^3$ shell which seems to leave no room for orbital degree of freedom. Owing to the atomic multiplet effects and some mixture of them by the small distortion of the CrSe$_6$ octahedron, Cr$^{3+}$ ions may have a small orbital moment and thus contribute to a small SIA.

An exchange anisotropy was proposed to be a major source of the MAE for CrI$_3$~\cite{2D_Lado2017, Kim_PRL2019}. The strong SOC of the heavy I $5p$ orbitals affects the FM Cr-I-Cr superexchange by the spin-orientation dependent electron hoppings, and thus yields an exchange anisotropy.
In CrSbSe$_3$, the major intra-chain FM couplings arise from the Cr-Se-Cr superexchange, and the weak inter-chain FM couplings have the contributions from the bridging Sb atoms whose $5p$ orbitals are more expansive. As the Se $4p$ and Sb $5p$ orbitals get heavily involved in those FM superexchanges, their strong SOC effects could also yield an exchange anisotropy as in CrI$_3$~\cite{2D_Lado2017, Kim_PRL2019}.

\begin{figure}[t]
	\includegraphics[width=7cm]{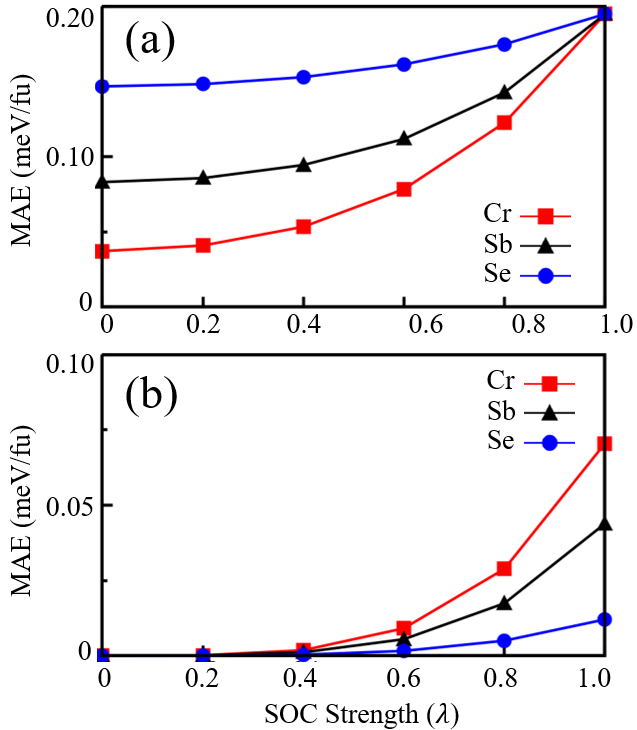}
	\caption {(a) Red line: evolution of the MAE (FM$^b-$FM$^a$) as a function of $\lambda_{\rm Cr}$ while keeping $\lambda_{\rm Sb}$ and $\lambda_{\rm Se}$ at 1 (their full values). Black (Blue) line: MAE as a function of $\lambda_{\rm Sb}$ ($\lambda_{\rm Se}$) while keeping two others at 1. (b) Red line: evolution of the MAE as a function of $\lambda_{\rm Cr}$ while setting $\lambda_{\rm Sb}$ and $\lambda_{\rm Se}$ at 0. Black (Blue) line: MAE as a function of $\lambda_{\rm Sb}$ ($\lambda_{\rm Se}$) while setting two others at 0.
	}
\label{fig7}
\end{figure}

We now identify each contribution of Cr, Sb, and Se elements to the MAE by performing LSDA+$U$+SOC calculations for the magnetization along the easy $a$ or hard $b$ axis. We first include the full SOC effect of the two elements, and tune the SOC effect of the third element from zero to its full value, see Fig.~\ref{fig7}(a).
The red, black, and blue curves mainly show the respective contribution of Cr, Sb, and Se to the MAE, albeit their mutual enhancement. The Cr contribution is about 0.158 meV per atom and mainly the SIA type, and Sb (Se) is about 0.112 (0.048) meV per atom and mainly the EA type. It seems that the SIA and EA have almost equal contributions to the MAE. Second, we turn off the SOC effect of the two elements, and tune the SOC effect of the third element from zero to its full value, see Fig.~\ref{fig7}(b). We see the Cr contribution of about 0.070 meV per atom to the SIA, and Sb (Se) contributions of 0.044 (0.012) meV per atom to EA, all of which are smaller than the corresponding contributions in the first case with a mutual enhancement of the SOC effects. These two sets of the results show that the SIA contribution from Cr and the EA one from Sb plus Se, are both moderate and each contributes about half of the MAE.

\begin{figure}[t]
	\includegraphics[width=8cm]{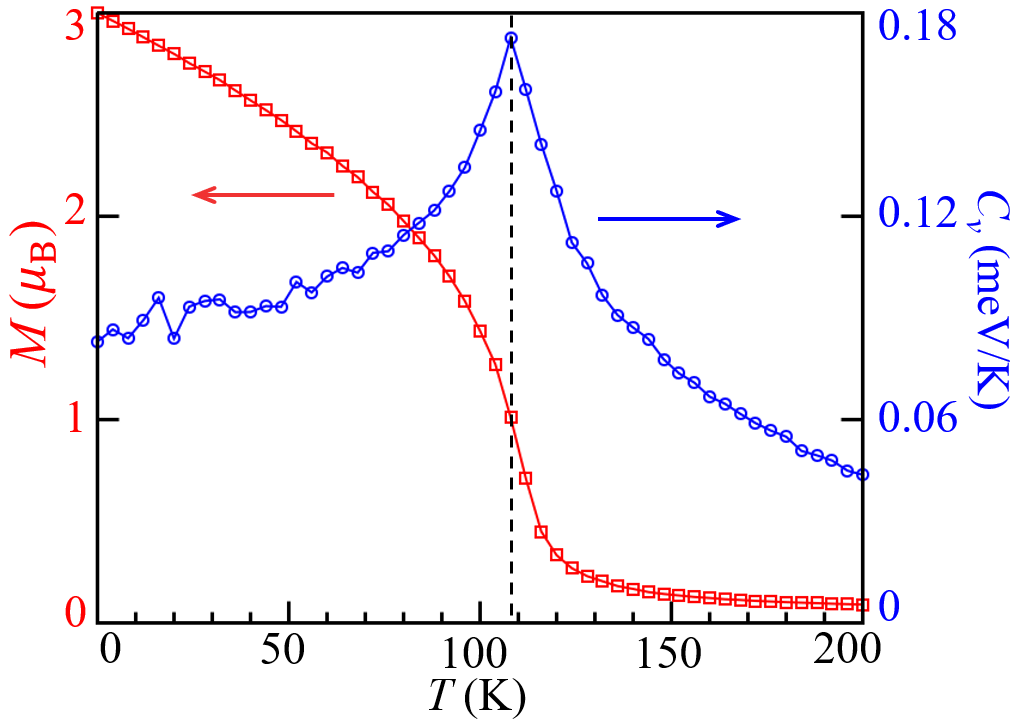}
	\caption {Monte Carlo simulations of the magnetization and magnetic specific heat of CrSbSe$_3$ as a function of temperature. $T_{\rm C}$ is estimated to be 108 K.
	}
\label{fig8}
\end{figure}

We now carry out Monte Carlo simulations to estimate the Curie temperature of the pseudo-1D FM semiconductor CrSbSe$_3$, using the above parameters of the magnetic exchange and anisotropy. The spin Hamiltonian is
\begin{small}
\begin{equation*}
\begin{aligned}
H = \sum_{\emph{k}} \sum_{\emph{i,j}} \frac{\emph{J}_{\emph{k}}}{2} \overrightarrow{S_{i}} \cdot \overrightarrow{S_{j}} &+  \sum_{i}\lbrace D(S_{i}^{c})^{2} + E_{n} [(S_{i}^{a})^{2} - (S_{i}^{b})^{2}] \rbrace \\
 = \sum_{k} \sum_{i,j} \frac{J_{k}}{2} \overrightarrow{S_{i}} \cdot \overrightarrow{S_{j}} &+ \sum_{i} D |\overrightarrow{S_{i}}|^{2} cos^{2} \theta \\
 &+ \sum_{i} E_{n} [ |\overrightarrow{S_{i}}|^{2} sin^{2} \theta (cos^{2} \phi - sin^{2} \phi) ],
\end{aligned}
\label{equation1}
\end{equation*}
\end{small}
where the first term stands for the isotropic Heisenberg exchange, the second term with $D$ describes the longitudinal ($c$ axis) anisotropy, and the last term with $E_n$ represents the transverse ($ab$ plane) anisotropy. The sum over $i$ runs over all Cr$^{3+}$ sites with $S$=$3/2$ in the spin lattice, and $j$ over the neighbors of each $i$ with their FM couplings $J_k$ ($k$=1-4, see Fig.~\ref{fig4} and Table~\ref{tb1}).
$D=-0.033$ meV and $E_n=-0.042$ meV are calculated using the above MAE values in the FM ground state (see Fig.~\ref{fig6}).
The Metropolis method~\cite{MC_Metropolis} and a $10 \times 10 \times 10$ spin lattice with the periodic boundary condition are used in our simulations. During the simulation steps, each spin is rotated randomly in all directions.
At each temperature, we use 10$^{8}$ Monte Carlo steps per site to reach an equilibrium, and 1.2$\times$10$^{8}$ steps per site for statistical averaging.
The magnetic specific heat, at a given temperature, is calculated according to
$C_v$=$( \langle E^2 \rangle - \langle E \rangle^2) / (k_{\rm {B}} T^2)$,
where $E$ is the total energy of the spin system.
The calculated magnetization and magnetic specific heat show that the Curie temperature
is 108 K, see Fig.~\ref{fig8}. Compared with the experimental $T_{\rm C}$ of 71 K, the present Monte Carlo simulations give an overestimated but reasonable $T_{\rm C}$, without considering the phonon contribution and spin-phonon interaction. This is often the case in low-dimensional magnetic materials~\cite{chens_2020, guoyilv_2018}.
Overall, the present work reasonably accounts for the band gap and magnetic properties of this pseudo-1D FM semiconductor.

\section{CONCLUSIONS}
In summary, we study the electronic structure and magnetic properties of the pseudo-1D FM semiconductor CrSbSe$_{3}$, using density functional calculations, superexchange model analyses, and Monte Carlo simulations.
Our results show that CrSbSe$_3$ is a strongly covalent Mott type semiconductor with a band gap of 0.65 eV and the total spin moment of 3 $\mu_{\rm B}$/fu (from the formal Cr$^{3+}$ $S$=3/2 state), both of which well reproduce the experiments.
We find two major FM couplings within the zig-zag spin chain along the $b$ axis, which are rationalized by the FM superexchange picture of the near-90$^\circ$ Cr-Se-Cr bonds, but the inter-chain FM couplings are one order of magnitude weaker, thus corroborating the pseudo-1D structure of CrSbSe$_3$.
Moreover, our calculations reproduce the experimental easy magnetization $a$ axis and the hard $b$ axis, and we find that the single ion anisotropy from Cr and the exchange anisotropy from Sb plus Se have nearly equal contributions to the calculated magnetic anisotropic energy of 0.19 meV/Cr. Using the parameters of the magnetic exchange and anisotropy, our Monte Carlo simulations give the Curie temperature of 108 K. Overall, this work reasonably accounts for the magnetic properties of the pseudo-1D FM semiconductor CrSbSe$_{3}$.

$Note$ $added.$ After submission of our paper, we noticed a paper \cite{note} which presents a theoretical study of the ferromagnetism for the  hypothetical 1D spin chain of CrSbSe$_3$. Their results of the  strong intrachain FM couplings are similar to ours for the real bulk material.

\section{ACKNOWLEDGMENTS}
This work was supported by National Natural Science Foundation of China
(Grants No. 12174062 and No. 12104307) and by the National Key Research and Development Program of China (Grant No. 2016YFA0300700).
\bibliography{CrSbSe3_PRMater}
\end{document}